\documentclass[12pt, letterpaper]{article}

\usepackage{graphicx}
\usepackage{amssymb}
\usepackage{amsmath}
\usepackage[figuresright]{rotating}
\usepackage{bm}
\usepackage[utf8]{inputenc}
\usepackage{mathtools}
\usepackage{cite}
\usepackage{algorithm}
\usepackage{algorithmicx}
\usepackage[noend]{algpseudocode}

\title{\fontsize{14}{15}\bfseries A\underline{$\mathcal{D}$}VE\underline{$\mathcal{R}$}SARIA\underline{$\mathcal{L}$}uscator: An Adversarial-DRL based Obfuscator and Metamorphic Malware Swarm Generator}

\author{
Mohit Sewak\\
\texttt{Microsoft R\&D, India}\\
\texttt{mohit.sewak@microsoft.com}
\and
Sanjay K. Sahay, Hemant Rathore\\
\texttt{BITS Pilani, Goa, India}\\
\texttt{\{ssahay, hemantr\}@goa.bits-pilani.ac.in}
}

\date{} 

\begin{document}
\maketitle

\begin{abstract}
Advanced metamorphic malware and ransomware, by using obfuscation, could alter their internal structure with every attack. If such malware could intrude even into any of the IoT network, then even if the original malware instance get detected, by that time it can still infect the entire network. The IoT era also required Industry 4.0 grade AI based defense against such advanced malware. But AI algorithm need a lot of training data, and it is challenging to obtain training data for such evasive malware. Therefore, in this paper, we present ADVERSARIALuscator, a novel system that uses specialized (adversarial) deep reinforcement learning to obfuscate malware at the opcode level and create multiple metamorphic instances of the same. To the best of our knowledge, A\underline{$\mathcal{D}$}VE\underline{$\mathcal{R}$}SARIA\underline{$\mathcal{L}$}uscator is the first-ever system that adopts the Markov Decision Process based approach to convert and find a solution to the problem of creating individual obfuscations at the opcode level. This is important as the machine language level is the least at which functionality could be preserved so as to mimic an actual attack effectively. A\underline{$\mathcal{D}$}VE\underline{$\mathcal{R}$}SARIA\underline{$\mathcal{L}$}uscator is also the first-ever system to use efficient continuous action control capable deep reinforcement learning agents like the Proximal Policy Optimization in the area of cyber security. Experimental results indicate that A\underline{$\mathcal{D}$}VE\underline{$\mathcal{R}$}SARIA\underline{$\mathcal{L}$}uscator could raise the metamorphic probability of a corpus of malware by $\ge 0.45$. Additionally, more than $33\%$ of metamorphic instances generated by A\underline{$\mathcal{D}$}VE\underline{$\mathcal{R}$}SARIA\underline{$\mathcal{L}$}uscator were able to evade even the most potent IDS and penetrate the target system, even when the defending IDS could detect the original malware instance. Hence A\underline{$\mathcal{D}$}VE\underline{$\mathcal{R}$}SARIA\underline{$\mathcal{L}$}uscator could be used to generate data representative of a swarm of very potent and coordinated AI based metamorphic malware attack. The so generated data and simulations could be used to bolster the defenses of an IDS against an actual AI based metamorphic attack from advanced malware and ransomware.
\end{abstract}

\section{Introduction}\label{introduction}

Industry 4.0 aims to proliferate networks of edge devices and autonomous control. Such a scenario is to the greatest advantage of the malware developers, especially the developers of metamorphic malware. Metamorphic malware are the second generation, advanced malware that could alter their internal structure using techniques like \textit{obfuscation} after each attack \cite{ye2017survey}. Even if the original malware instance is known or detected a while later, its still challenging to detect, locate and quarantine all its metamorphic instances of the malware in the network. Threat scenario becomes even more challenging if the metamorphic instances are from different original malware or a cross between other known malware. 
\par
For any traditional, or even classical Machine Learning (ML) based Intrusion Detection Systems (IDS) \cite{rathore2020identification, rathore2020detection}, even a single such metamorphic attack scenario is akin to an extreme threat scenario arising from \textit{multiple simultaneous zero-day attack} and has the potential of compromising the entire network and the \textit{Factory/Industry} setup. Though it may be possible to use advanced Deep Learning (DL) based IDS \cite{SewakSNPD, sewak2020overview}, but such complex DL based IDS need a lot of labeled data for training before they can detect the obfuscated instance generated by a metamorphic malware. Though, there exist synthetic adversarial data generators that could generate perturbations at the binary level that can evade a DL based IDS \cite{Adversarial-DL-Binaries, rathore2020robust, drl-evading-botnet, MalGAN, IDSGAN}; but training against the data generated by such systems could potentially cause more harm to the IDS instead of \textit{incrementally} training them against metamorphic instances of known malware \cite{MalGAN}. This is because the lowest level at which an actual code level obfuscations could be identically represented and the file functionality could be identify-ably preserved is at the assembly language (sequence of opcodes) level. But owing to the length and the possible combination of opcodes that makes a unique malware, to simulate a metamorphic malware generator even at the opcode level requires solving a very complex Markov Decision Process (MDP) with very high cardinality action-space.
To the best of our knowledge, there exists no system that claims to generate AI driven obfuscations at the opcode level.
\par
Therefore, in this paper we present a novel system named \textbf{ADVERSARIALuscator}, for \textit{\textbf{Adversarial} \underline{\textbf{D}}eep \underline{\textbf{R}}einforcement \underline{\textbf{L}}earning based obf\textbf{uscator} and Metamorphic Malware Swarm Generator}, that could solve such a complex MDP to generate swarms of metamorphic malware obfuscated at the opcode level.
Earlier in our initial work named DOOM \cite{sewak-ubicomp-doom} we solved this MDP which has limited unique DRL agents that could not be used to propagate a swarm attack.
On the other hand, A\underline{$\mathcal{D}$}VE\underline{$\mathcal{R}$}SARIA\underline{$\mathcal{L}$}uscator is a matured, multi-agent implementation of a complete metamorphic swarm generation system.
The metamorphic malware swarm data generated by A\underline{$\mathcal{D}$}VE\underline{$\mathcal{R}$}SARIA\underline{$\mathcal{L}$}uscator and the simulations obtained from the MDP environment we created can help bolster defenses of any (even networked) IDS against metamorphic malware swarm attacks. The A\underline{$\mathcal{D}$}VE\underline{$\mathcal{R}$}SARIA\underline{$\mathcal{L}$}uscator system can not only be used for on online IDS, but could also scan files, directories and databases.

\par The remaining of the paper is organized as follows. In section \ref{sec:motivation} we cover the background and motivation, whereas in section \ref{sec:background-related-work} we discuss the related work on several aspects of ADVERSARIALuscator.
Next, in section \ref{sec:architecture}, \ref{sec:rl-environment} and \ref{sec:rl-agent} we cover the details of the architecture, the custom RL environment and the DRL agent for ADVERSARIALuscator.
Then, in section \ref{sec:results} we describe the results of the various experiments carried out on the setup as described. We discuss the implications of the achieved results in section \ref{sec:discussion} and finally conclude the paper in section \ref{sec:conclusion}.

\section{Background and Motivation} \label{sec:motivation}
In this section we cover the background and motivation related to different aspects of obfuscation and ADVERSARIALuscator.

\subsection{Obfuscation vs. Binary Perturbations}\label{sec:ADVERSARIALuscator-advantages}
As opposed to generating features with perturbations of binary sequences in the network payload that could makes a malicious payload be classified as non-malicious, the A\underline{$\mathcal{D}$}VE\underline{$\mathcal{R}$}SARIA\underline{$\mathcal{L}$}uscator system creates obfuscations of actual malware at the opcode level. Doing so the A\underline{$\mathcal{D}$}VE\underline{$\mathcal{R}$}SARIA\underline{$\mathcal{L}$}uscator system not only mimics an actual metamorphic malware (at feature vector level) which an existing system is more likely to actually encounter than random perturbations but is also compatible with Intrusion Detection Systems (IDS) that could also be used for malware detection at a file system level and provide many other benefits as we discuss next.

Instead of working at the binary sequence level, many advanced IDS, work at a more semantically informative opcode sequence level to be more effective in identifying such advanced threats. If sufficient training data is available, some IDS could even create defence against existing instances of Metamorphic malware as well \cite{SewakSNPD}. 

Since most of ML and DL techniques work on the implicit or explicit assumption of identifying patterns in a specific distribution of training data, therefore if the inference (new detection candidate samples) data distribution is significantly different from that of their training data, these techniques may not only provide an ambiguous classification, but in some cases may also provide an incorrect classification. In the context to an IDS this mean that the IDS may not only fail to detect the incoming malware as a threat, or at least as doubtful (and thus warranting additional static or dynamic analysis), but may even pass it as a non-malicious sample. Obfuscation could create such potent metamorphic malware samples which may seemingly come from a different (desirably non-malicious) distribution and may evade even most of the advanced ML/ DL based IDS.

The IDS' that work with binary-sequence based features, could be easily evaded by data generated by creating synthetic perturbations at a binary level \cite{Adversarial-DL-Binaries, rathore2020robust, drl-evading-botnet, MalGAN, IDSGAN}. 
Such generated perturbations/ noise added to a malicious binary data could create synthetic samples, which to an IDS seem to have come from a non-malicious binary network traffic payload and hence corrupt the IDS' detection. The intention behind such mechanism is that the samples that evade the IDS may be used to re-train the IDS to augment its defenses against a new malware or a new instance (supposedly obfuscated) of an existing malware. The underlying implicit assumption being that some of the samples created by these perturbations may have features that are identical to that of an actual new malware or an obfuscation of an existing malware, and hence re-training with this additional data will have a monotonically desirable impact on the performance (definitely increase or at least will not deteriorate the performance) of the IDS.

These implicit assumptions may not only prove to be unfounded sometimes, but may even be counter intuitive at other times. This is because: 
\begin{itemize}
    \item As indicated in \cite{MalGAN} these methods, especially the ones based on gradient-attack based techniques (like GANs) make re-training of IDS ineffective.
    \item Such generated perturbations may not represent an actual opcode/ instruction and their insertion point in the binary sequence may not correspond to a logical start or a logical end of an opcode. Such mechanism generate synthetic binary sequences and insert them in a manner that may not represent a similar one generated from the compiler for an original file (malicious or non-malicious). Hence despite claims of functionality-preservation, synthetic data generated by such systems may never actually represent an actual malware that a developer would create and which can infect a system. 
    \item Moreover since each executing processor architecture understand different set of opcode and each opcode may produce a very unique set of binary sequence, so despite claims in the respective papers of functionality-preservation, the random sequence of binary data produced may not even be comprehend-able by the target executing process architecture.
    \item Lastly when the IDS is trained with so much data from a distribution that may not represent features of a likely obfuscated malware that the IDS would probably face in real deployment, it  may even make the IDS over-fit \cite{Anderson_PE_RL} or reduce the proportion of actual malware sample data from its training, thus reducing its effectiveness against real new inference data.
\end{itemize} 

% In either of the above cases, the targeted binary noise that evades the IDS may not only be un-suitable for improving the IDS's response against (undetectable) metamorphic instances of existing (detectable) malware variants, but may even deteriorate their likely performance against some of the actual instances of the metamorphic malware that they may encounter later. 
% At the binary level it is difficult to ascertain if a specific system generated sequence of binary represent an actual compiler generated and processor compatible representation of a malware. Therefore such synthetic training data even if essentially required to retrain an IDS, should best be created at a (human understandable) code or at least at an instruction/ opcode level.

Another critical drawback with the entire approach is that it is entirely dependant on retraining IDS's classifier with synthetic data for improving the IDS's detection capabilities against (new or) metamorphic malware. Whereas a more potent technique to enable the IDS to detect the obfuscated malware could be to de-obfuscate \cite{deobfuscation-windows-api-call} the malware before sending it (or its extracted features) to the IDS's classifier. Classifies if just one of the sub-component of the IDS. The other sub-components being the feature extractor and optionally a feature transformer, pre-processor etc. As indicated earlier, in some cases it may be counter-intuitive to re-train a classifier on synthetic data; therefore sub-components or sub-systems could be created/ trained that could try to solve the problem of metamorphic malware detection without mandating re-training of the IDS's classifier. 

\subsection{Uses of Obfuscations created by ADVERSARIALuscator}\label{sec:uses-obfuscations}
The A\underline{$\mathcal{D}$}VE\underline{$\mathcal{R}$}SARIA\underline{$\mathcal{L}$}uscator system creates metamorphic malware features at the opcode level. One additional benefit of creating obfuscations at this level is that these could be used not only to re-train the IDS's classifier, but may also be used to augment other sub-components of the IDS as well (like the feature extractor or pre-processor), thus alleviating the need to tamper with a classifier that is well trained on actual malware data distribution. Also ancillary (internal or external to IDS) sub-systems could be created that could de-obfuscate the obfuscation of a metamorphic malware to normalize it to its (existing) base malware variant before sending it to the IDS's classifier for detection. 

Creating a normalized representation of the metamorphic instance of specific malware variant is not only important from a cyber-forensic perspective, but is also critically required for some advanced multi-nominal IDS and other similar systems that also detect the variant of the malware (besides class) for further action. Therefore the opcode level obfuscations generated by the A\underline{$\mathcal{D}$}VE\underline{$\mathcal{R}$}SARIA\underline{$\mathcal{L}$}uscator system could be used in either of these ways:

\begin{itemize}
    \item For improving the IDS's classifier detection by re-training it with training data additionally augmented with the obfuscated metamorphic instances of existing malware.
    \item For training/ augmenting other internal sub-systems of the IDS with capabilities that could de-obfuscate the incoming file's extracted features before sending it to IDS's classifier.
    \item For creating/ training other external sub-systems for normalizing obfuscations of different variant of existing malware to augment an existing IDS without mandating the IDS to be even aware of their existence. 
\end{itemize}

% But to alleviate all the drawbacks associated with binary perturbation techniques and to enable all these benefits associated with obfuscation based techniques, we need to create metamorphic obfuscation of existing malware at an instruction/ opcode or higher (code) level. Generating Obfuscations at an opcode (or higher) level is not a new idea. In fact any perturbation at a binary level could not even be called an obfuscation, as this terminology is restricted to activities (like junk code insertion, junk instruction insertion etc.) that could effectively be carried out only at a higher abstraction levels (opcode/ instructions or code). So binary perturbations based systems cannot even technically create a metamorphic or any obfuscated malware, leave aside creating defense against one. 

\section{Related Work}\label{sec:background-related-work}

In this section we cover the related work on different aspects of obfuscation and ADVERSARIALuscator.

\subsection{DRL-Agent as IDS Adversary}\label{sec:GAN}
One reason why most of the existing perturbation-creation based systems still work at binary level is that many of these systems work on the basis of adversarial machine learning systems like the Generative Adversarial Networks (GANs) \cite{MalGAN, IDSGAN, rathore2020robust, UsamaGAN}. In such an adversarial setting, there exists a `Discriminator Network' ($\mathcal{D}$) which mimics an IDS's classifier and tries to detect any new sample created by another `Generator Network' ($\mathcal{G}$) of the system. In a typical GAN both the $\mathcal{D}$ and $\mathcal{G}$ networks are comprised of Convolutional Neural Networks (CNNs) \cite{Sewak-CNN}, and it is intuitive to transform a binary network traffic as a CNN or other similar input layer so it makes it easier to work at binary sequence level with GAN based adversarial techniques. 

Some advanced systems in this category could use Reinforcement Learning (RL) \cite{Anderson_PE_RL} or even Deep Reinforcement Learning (DRL) \cite{Sewak-DRL} agents \cite{drl-evading-botnet} to replace the $\mathcal{G}$ network. But even these Adversarial RL/DRL based systems also work at the binary network traffic level. The reason for even these to to work at the binary level could be that most of the popular RL algorithms (like Q Learning) or even DRL algorithms (like Deep Q Networks) could find efficient solutions to only low cardinality action space based RL tasks. Therefore these systems could only learn a policy comprising of just a few selected actions that the agent could take. These actions could be like adding or removing a very specific sequence of binary instructions. In most of such systems the sequence of binaries that could be inserted/ deleted is also either very abstract or otherwise restricted. This further adds bias to the created data even if the resulting binary sequences remains operation after such alterations.

% an opcode or human understandable (higher) code level could change not only the distribution of the binary-sequences of the file, but also the opcode frequency or even sequence enough to even evade many ML, or DL based advanced IDS.

 But training an Adversarial DRL/RL based agent for this purpose is not a trivial task. The instruction-set for any architecture may contain thousands of opcodes. Therefore the `Markov Decision Process' (MDP) that could mimic the obfuscation of a malware holistically at the sophistication level represented by inserting any number of junk instructions in different combination represents a very high cardinality observation and action space. Solving an MDP require a RL agent that could find efficient solution to such RL task. Hence popular RL or DRL algorithms like the ones employed in the existing literature in this field are not suitable to learn such obfuscation task at the required sophistication level.
 
 The A\underline{$\mathcal{D}$}VE\underline{$\mathcal{R}$}SARIA\underline{$\mathcal{L}$}uscator system hence has to use specialized DRL agents to learn this RL task and create the desired obfuscations. Experimental results indicate that the obfuscated malware created by A\underline{$\mathcal{D}$}VE\underline{$\mathcal{R}$}SARIA\underline{$\mathcal{L}$}uscator could evade (classified as non-malicious or ambiguous) even the most potent malware intrusion detection systems and hence could mimic even an extreme `zero day attack' from multiple simultaneous metamorphic malware. To the best of our information, A\underline{$\mathcal{D}$}VE\underline{$\mathcal{R}$}SARIA\underline{$\mathcal{L}$}uscator is the first system that takes a Markov Decision Process based approach to convert and find a solution to the problem of opcode level obfuscation, and it is also the first ever system to use continuous action control capable deep reinforcement learning agents the like Proximal Policy Optimization (PPO) in the area of malware generation and defense.

The PPO agents acts as (multiple) $\mathcal{G}$ networks in the adversarial setup. For enacting as the $\mathcal{D}$ networks, we use the complete IDS (including pre-processing, feature-selection and transformation) that claimed to provide the most superior performance (accuracy of 99.21\%with an FPR of 0.19\%) \cite{SewakSNPD} over a standardized malware dataset with mixed types and generation of malware. We also use the associated malware data on which this performance was achieved \cite{Malicia} to train ADVERSARIALuscator.

% \par The main contributions of this paper are:
% \begin{itemize}
%     \item Development of a \textbf{Policy-Optimization} based (PPO) Deep Reinforcement Learning based Obfuscation System that could obfuscate any file at the opcode level
%     \item Underlying \textbf{safeOpCodeObfuscation} mechanism that ensure that the generated obfuscation could be only used to generate data to train defence solutions and not to perpetrate an advanced attack (without significant reverse engineering).
%     \item Development of Deep Reinforcement Learning \textbf{Environment} that is generic enough to train multiple types of off-the-shelf DRL agents.
% \end{itemize}

\subsection{Code/opcode level obfuscation} \label{related-work}
There have been many attempts to generate obfuscations at the code level
 \cite{metamorphic_obfuscation_Borello2008}, but these are not scalable. Later efforts were also made to use machine learning models  \cite{metamorphic_virus_generator_desai} to automate to some extent the obfuscation mechanism. Most of these methods does not replicate the advance multiple-simultaneous botnet metamorphic attacks required to train a deep reinforcement learning adversary. There has been attempt to use CNN based Generative Adversarial Networks (GANs) \cite{rathore2020robust, MalGAN, IDSGAN, UsamaGAN} as well but these constitutes mostly using supervised training approaches on specific algorithms. Recently Deep Reinforcement Learning (especially Q Learning) has been utilized \cite{drl-evading-botnet} to alter the binary code of the file to evade attacks. Such systems are not only limited to very small action-space problems (here limited to adding some specific 4-bit code) but even these could not be used to mimic an actual obfuscations (which require a code/opcode level treatment).
 To the best of our knowledge our system is the first ever to work directly on the numerous of opcodes (hence requires solving a large-action space reinforcement problem) to mimic obfuscations that could be used generate data to train any solution (even deep learning or reinforcement learning based) that works on opcode frequency features for malware detection.

\section{Architecture of the `ADVERSARIALuscator'}\label{sec:architecture}

\begin{figure}[htbp!]
    \centering
    \includegraphics[width=0.97\linewidth]{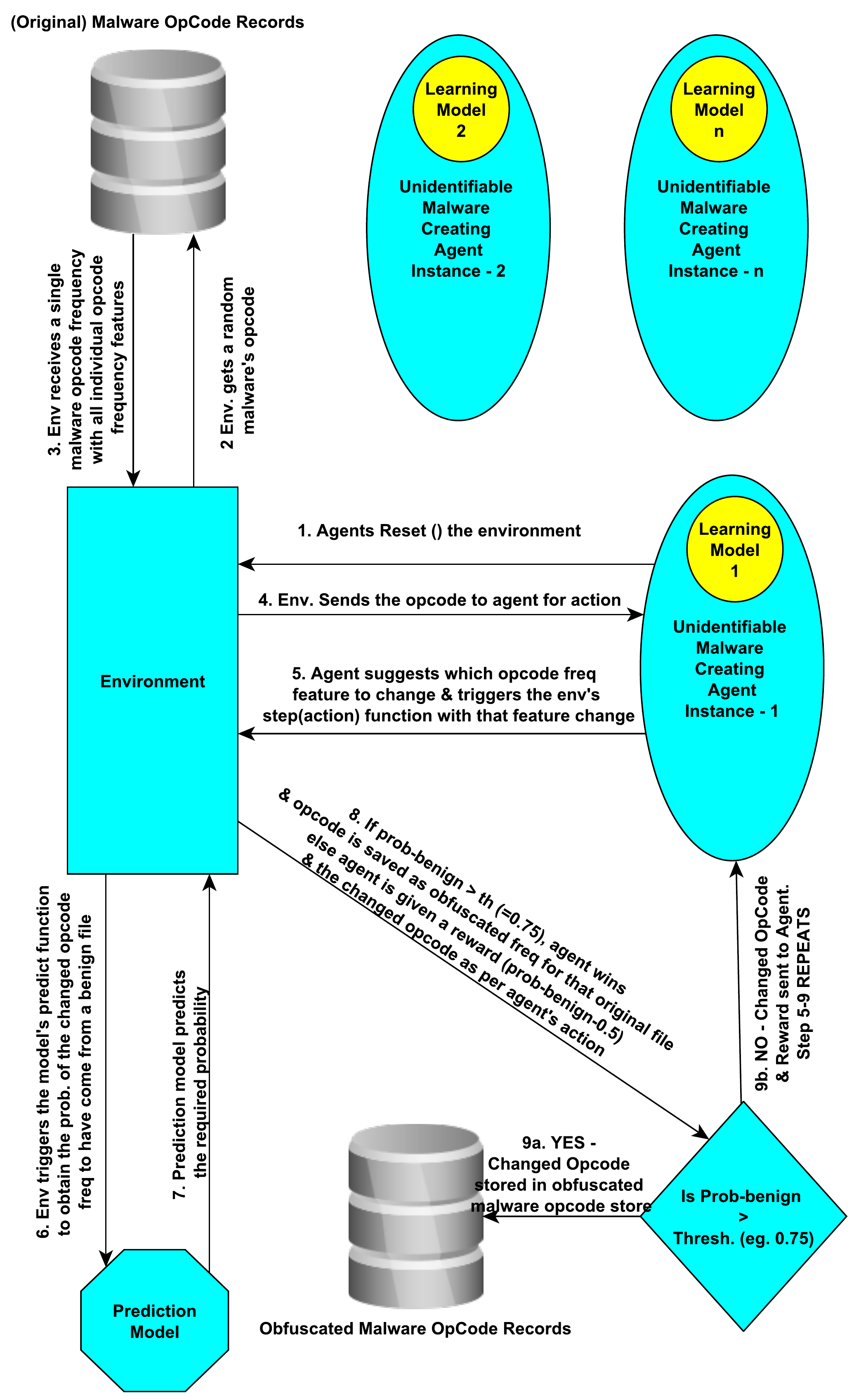}
    \caption{Agent Training Process Flow}
    \label{fig:process_flow}
\end{figure}

The architecture of the \textit{ADVERSARIALuscator System} is shown in figure \ref{fig:process_flow}. This architecture broadly consists of four subsystems:
\begin{enumerate}
    \item The opcode repository of the original malware files (collected from Malicia dataset \cite{Malicia}) and its subsequent obfuscated instances (as produced by the the A\underline{$\mathcal{D}$}VE\underline{$\mathcal{R}$}SARIA\underline{$\mathcal{L}$}uscator system) and corresponding metadata (like original malware mapping, similarity scores with normalized malware feature vector etc.).
    \item Repository of existing trained IDS to act as the adversary (`Discriminant Network' as in GAN).
    \item A custom RL environment (to emulate the MDP for agents to learn against).
    \item The Obfuscating DRL Agent(s).
\end{enumerate}

% The process flow for the design of the complete system is as shown in figure \ref{fig:process_flow} below. It has broadly 3 sub-systems, namely the (malicious) opcode extraction and repository, an trained and effective opcode frequency transformation and classification system, a custom malware obfuscation training reinforcement learning environment, and a Deep Reinforcement Learning (DRL) agent compatible with both the custom environment and also the type of reinforcement learning problem at hand (conceptually and mathematically).

In figure \ref{fig:process_flow}, the DRL agent interacts with the environment to train against \textit{episodic} tasks which can alter the presented opcode frequency. Such alterations could be strictly additive corresponding to a net addition of opcode instructions thus mimicking junk instruction/ code insertion for obfuscation \cite{ObfuscationTechniques}. The agent may even choose choose to insert opcodes which though are present in the instruction set of the architecture but not present in the file. 

During the initialization of an \textit{episode} the environment fetches a randomly selected malware variant's opcode frequency feature vector from the malware feature vector repository. Then the environment scores the obtained feature vector using the available IDS. The IDS acts as an adversary and provides the the (initial) non-malicious (prediction) probability ($\mathbb{P_\text{non-malicious}}$) of the selected malware variant. This is stored in the metadata to compute the reduction in $\mathbb{P_\text{non-malicious}}$ in each episode by the corresponding agent. The environment then serves this feature vector (\textit{called observation as in RL}) to the agent to propose an action. The action proposed by the agent is the sequence index of the opcode whose frequency it proposes to be increased, thus representing a real-life action of \textit{inserting new opcodes/ instructions to the existing malware}.

The environment in each training step subsequently invokes the available IDS to obtain the probability of the generated opcode frequency vector (modified as per the agent's `action') to have come from a `Non-Detectable Malware File' \textbf{(NDMF)} file (thus representing a successful obfuscation of malware to non-malicious) ($\mathbb{P}_{\text{NDMF}} = 1-\mathbb{P}_{\text{DMF}}$, where \textbf{DMF} stands for `Detectable Malware File'). A decent increment in $\mathbb{P}_{\text{NDMF}}$ from the initial state (most of the malicious files as used in this system had a $\mathbb{P}_{\text{DMF}} \approx 1.0$, ($\lim_{\mathbb{P}_{\text{DMF}} \to 0}$)  by the IDS to a level where the IDS's classifier could not disambiguate the file instance is considered as the first success criteria. The second success criteria being taking the $\mathbb{P}_{\text{DMF}}$ to the other extreme so that the IDS starts classifying the resulting feature vector as to come from a non-malicious file with very high probability. The DRL agents trains over multiple such training episodes to update and refine an action-policy to \textit{obfuscate} any malware variant to achieve both or at-least the first success criteria. 
\raggedbottom

\textit{Multiple} such DRL agents could be instantiated and trained with varying degree of dissimilarity from the other DRL agents so that each could learn to obfuscate a given malware using a slightly different action-policy and subsequently producing multiple dissimilar metamorphic instances of the same malware variant. There are different methods to introduce such dissimilarities. Some of these methods could range from one extreme of changing the complete underlying algorithm of the DRL agent, to other extreme where a the dissimilarities could be achieved by simply changing random number seed of various instances of the same DRL agent. The system thus creates the required collection of multiple metamorphic instances that could mimic a very powerful and sophisticated attack on high value assets like defence installations, nuclear plants, strategic financial systems, and government assets. Therefore such created obfuscations could represent an ideal dataset to be used for creating a defensive mechanism and solution to avert such sophisticated attacks on similar high value assets.

\subsection{safeOpcodeObfuscation Mechanism} \label{sec:safety-mechanism}
There are many malware creation tool-kits available in public domain, which are already being used to create very potent malware variants. Such tool-kits do not use advanced AI to create virtually undetectable obfuscation, whereas the A\underline{$\mathcal{D}$}VE\underline{$\mathcal{R}$}SARIA\underline{$\mathcal{L}$}uscator system is capable of generating such sophisticated obfuscations virtually undetectable even with multiple ML/ DL detection systems. Hence the A\underline{$\mathcal{D}$}VE\underline{$\mathcal{R}$}SARIA\underline{$\mathcal{L}$}uscator system or an implementation of the same in the wrong hand could potentially create huge negative impact. Therefore to obviate such actions we have purposely designed the process such that the obfuscation part of the system would work only at the opcode level and could not be used to create a malicious executable. 

Despite it could have been possible to use the extracted opcode sequence as-is instead of frequency vector. By opting for the later we have kept the obfuscation process purposely \textit{intractable} in this system such that it works on abstraction of extracted opcode frequencies (instead of actual opcode sequences). If the original opcode sequence is not archived then the frequencies once extracted could not be re-sequenced to create a valid assembly (.asm) file so as to generated an malicious executable. 

Since  the objective of the A\underline{$\mathcal{D}$}VE\underline{$\mathcal{R}$}SARIA\underline{$\mathcal{L}$}uscator system is to strengthen the cyber-defense, we have taken the extra precautions that A\underline{$\mathcal{D}$}VE\underline{$\mathcal{R}$}SARIA\underline{$\mathcal{L}$}uscator (at least without major post-processing or reconfiguration) could not be used by cyber-attack-perpetrators to defeat the very objective and use A\underline{$\mathcal{D}$}VE\underline{$\mathcal{R}$}SARIA\underline{$\mathcal{L}$}uscator to generate a flurry of multiple simultaneous `zero day' metamorphic malware attack which the existing IDS may not be able to defend against. 

\subsection{functionalityPreserving Metamorphosism} \label{sec:functionality-preserving}
As reasoned in section \ref{introduction}, most of the systems that work at binary level, despite their respective claims, cannot produce a metamorphic malware (or even claim to produce an obfuscated malware) that mimics an actual program compiled from a real compiler and that is compatible with the downstream execution architecture, or is even compatible with its instructions.

ADVERSARIALuscator instead works at the opcode level, identifies the malware at the instruction level and also create obfuscations at the instruction level. The functionality of a given program is indicated by the sequence of the instructions available in its assembly produced by the compiler. As explained in section \ref{sec:architecture}, A\underline{$\mathcal{D}$}VE\underline{$\mathcal{R}$}SARIA\underline{$\mathcal{L}$}uscator only inserts junk instructions and does not remove the existing to preserve the functionality even at the instruction level. Doing so A\underline{$\mathcal{D}$}VE\underline{$\mathcal{R}$}SARIA\underline{$\mathcal{L}$}uscator does not only preserves the intended functionality of the program, but even risks that it could be used to create an actual malware. Therefore despite the desired outcome of functionality preservation, we had to also the safety mechanism as described in section \ref{sec:safety-mechanism} to ensure that where for all the theoretical purposes of the claims and experiments the functionality is preserved, in practice it remains intractable.

\section{Custom Reinforcement Learning Environment}\label{sec:rl-environment}

The environment serves a major role in reinforcement learning. Here its role is to present a current-state of the opcode frequency vector of the malware file from the repository to the agent to act upon, and then give it an appropriate reward and the corresponding next-state. The next-state here represents the resultant opcode frequency vector after changing the specific opcode\'s frequency as suggested by the agent in the current step. The next-state for the current-step becomes the current-state for the next-step in the environment.

The \textit{current-state}, \textit{action}, \textit{reward}, \textit{next-state} \textbf{(S-A-R-S)} cycle continues until a terminal-state is reached (for an episodic task) or until a (parameterized) predefined number of steps are exhausted. On reaching such scenario, the environment resets and re-instantiates itself with the opcode frequency vector of a randomly chosen new malware file from the repository and resets other necessary variables like initial $\mathbb{P_{\text{DMF}}}$, turns\_completed, total\_episode\_reward, total\_discounted\_episode\_reward, is\_complete (current episode completion) flag as in the algorithm \ref{alg:reset}. From the `reset' to the episode completion, the environment responds in each step as given in algorithm \ref{alg:step}.

\subsection{The structure of `State'}\label{env-state}
The state in our experiment comprises of a vector of whole numbers corresponding to the set of opcode frequency for a given file. We use the same unique opcode set as used by Sewak et. al. \cite{SewakSNPD}. We also use the same IDS which they claimed to have produced the best performance and in their work. Their IDS claimed an accuracy of 99.21\%with an False Positive Rate of 0.19\% on the Malicia dataset \cite{Malicia} which is the best performance available in literature on any standardized malware dataset. This dataset along with the collected non-malicious files for the work resulted in a set of 1612 unique opcodes (instruction set). Correspondingly we have a state comprising of 1612 dimension `Box' Space with a  permissible range of [0, 10000] $ \in {\mathbb{Z}}^{1612} $. The `Box' Space is an `Open AI's `Gym' \cite{OpenAIGym} compatible API reference class for defining the state for a Reinforcement Learning environment such that it could be used against any standardized setup/ agent.

\subsection{The design of `Action'}\label{env-action}
The `action' in our reinforcement learning problem is the specific opcode index that needs to be altered in a given step of an episode. We used a large cardinality discrete action control capable agent. The agent determines which opcodes' frequency should be altered and in which direction it should be altered (increased or decreased). The magnitude of alteration could be a fixed constant or a trainable parameter.
In this approach we have $\mathcal{N}_{\text{observation}} \times 2$ actions, the first $\mathcal{N}$ actions corresponding to a an increase in the specific opcode frequency by a constant $\mathcal{C}_{\text{increment}}$, and the next $\mathcal{N}$ actions represent an act of decreasing the corresponding opcode frequency by a fixed amount $\mathcal{C}_{\text{decrement}}$, where $\mathcal{C}_{\text{increment}}, \mathcal{C}_{\text{decrement}} \in \mathbb{N}$. In our implementation we have kept $\mathcal{C}_{\text{increment}} = \mathcal{C}_{\text{decrement}} = 5$. Also since from the perspective of obfuscation, the easiest way of creating multiple obfuscation often increases the opcode frequency by adding junk code, instructions etc. \cite{ObfuscationTechniques}. Therefore, to mimic this effect we allow the agent's action only a \textit{net} increase in individual opcode frequency from their initial level (as in original malware). An action with a \textit{net} effect of decreasing an individual opcode frequency below its original level results in returning the the same state as before the action and a commensurate reward.

\subsection{The `Reward' function}\label{env-reward}
How and what the agent learns is to a considerable degree dependant upon the reward/ penalty criteria and the magnitude thereof. Our first objective is that the agent should learn to alter the opcode frequency so as to substantially enhance the probability that the opcode frequency feature vector belongs to an NDMF (considers an non-malicious) instead of to a DMF (malicious) file. 

For a binomial classifier the unbiased cut-off threshold for separating the two class is midway i.e. the 0.5 probability mark as in equation \ref{eq:prob-threshold}. That is  $\text{given: } \text{opcode} \in \mathbb{W}^{1612}: $
\begin{multline}
\mathbb{P} (\text{opcode}_\text{file} \mid file \subseteq \{\mbox{non-malicious files}\}) \\
= \mathbb{P} (\text{opcode}_\text{file} \mid file \subseteq \{\mbox{malicious files}\}) = 0.5
\label{eq:prob-threshold}
\end{multline}
Owing to the IDS' established performance, most of the malicious files' opcodes have a $\mathbb{P}_\text{\mbox{non-malicious}} \approx 0.0 $ to start with. We penalize any resulting opcode frequency feature vector that has predicted  $\mathbb{P}_{non-malicious} \le 0.5$ and reward the ones with $\mathbb{P}_{non-malicious} \geq 0.5$ proportionally. That is in each step the reward given to the agent is:

\text{Given:}
$\mathbb{P}_\text{non-malicious} = \mathbb{P} (\text{opcode}_\text{file} \mid file \subseteq \{non-malicious files\})$
We have reward: $\mathbb{R} = \mathbb{P}_\text{non-malicious} - 0.5 $.

But this reward criteria when in isolation has a drawback that it encourages long trajectories resulting in greater positive rewards instead of favoring a shorter trajectory that quickly reaches a very high $\mathbb{P}_\text{non-malicious}$. One way to ensure that distant rewards are not favored much is to increase the `discounting-factor' ($\gamma$). But since the $\gamma$ could only be altered by agent's algorithm and is not in the scope of the environment's, so the reward mechanism cannot use lowering the discounting-factor enough to ensure that high instantaneous-rewards are more favorable
than lower cumulative-discounted-rewards. So to overcome this drawback, we have another (instantaneous) reward component. This reward component is given by the environment to the agent (in addition to the one stated above), when the agent to manage to alter the opcode frequency enough such that the file is almost unambiguously classified as non-malicious (second objective). This reward is high enough to easily surpass even multiple cumulative (even discounting given $\gamma < 1)$ rewards. This criteria occurs when the $\mathbb{P}_\text{non-malicious} \geq \mathbb{P}_\text{threshold}$. Where, $\mathbb{P}_\text{threshold}$ is a high/ threshold probability of non-malicious (say 0.90). Therefore, as a step function now the reward can be given as equation \ref{eq:reward_function} below.
\begin{equation}
    \text{reward}= 
    \begin{cases}
        \mathbb{P}_\text{non-malicious} - 0.5,& \text{if } \mathbb{P}_\text{non-malicious}\\ &\le \mathbb{P}_\text{threshold}\\
        \mathcal{R}_\text{goal}, & \text{otherwise} 
    \end{cases}
    \label{eq:reward_function}
\end{equation}
Where $\mathcal{R}_\text{goal}$ could either be constant or a variable dependant upon the maximum steps allowed in the episode. Each new episode starts from the reset of environment. On reset, a random malware file's opcode is retrieved and the episode continues until the goal is reached (achieving high non-malicious probability threshold) or exhausting the maximum permissible steps for the particular episode. Here we set $ \mathcal{R}_\text{goal} = $ Max\_Permissible\_Steps\_in\_an\_episode so that we could balance the needs for setups that have larger episode lengths to allow for slow convergence of complex agents with too many optimize-able parameters and have max\_permissible\_step adaptive $\mathcal{R}_\text{goal}$ such that it is always greater than any cumulative reward over even a long episode. Alternatively, or additionally, we could give a large penalty (negative reward) if the max\_permissible\_steps is exhausted without the agent managing a $ \mathcal{R}_\text{goal}$. We did not use this second mechanism in our experiment.

\subsection{The custom `ObfuscationEnvironment' design and algorithm }
\begin{figure}
    \centering
    \includegraphics[width=3.3in]{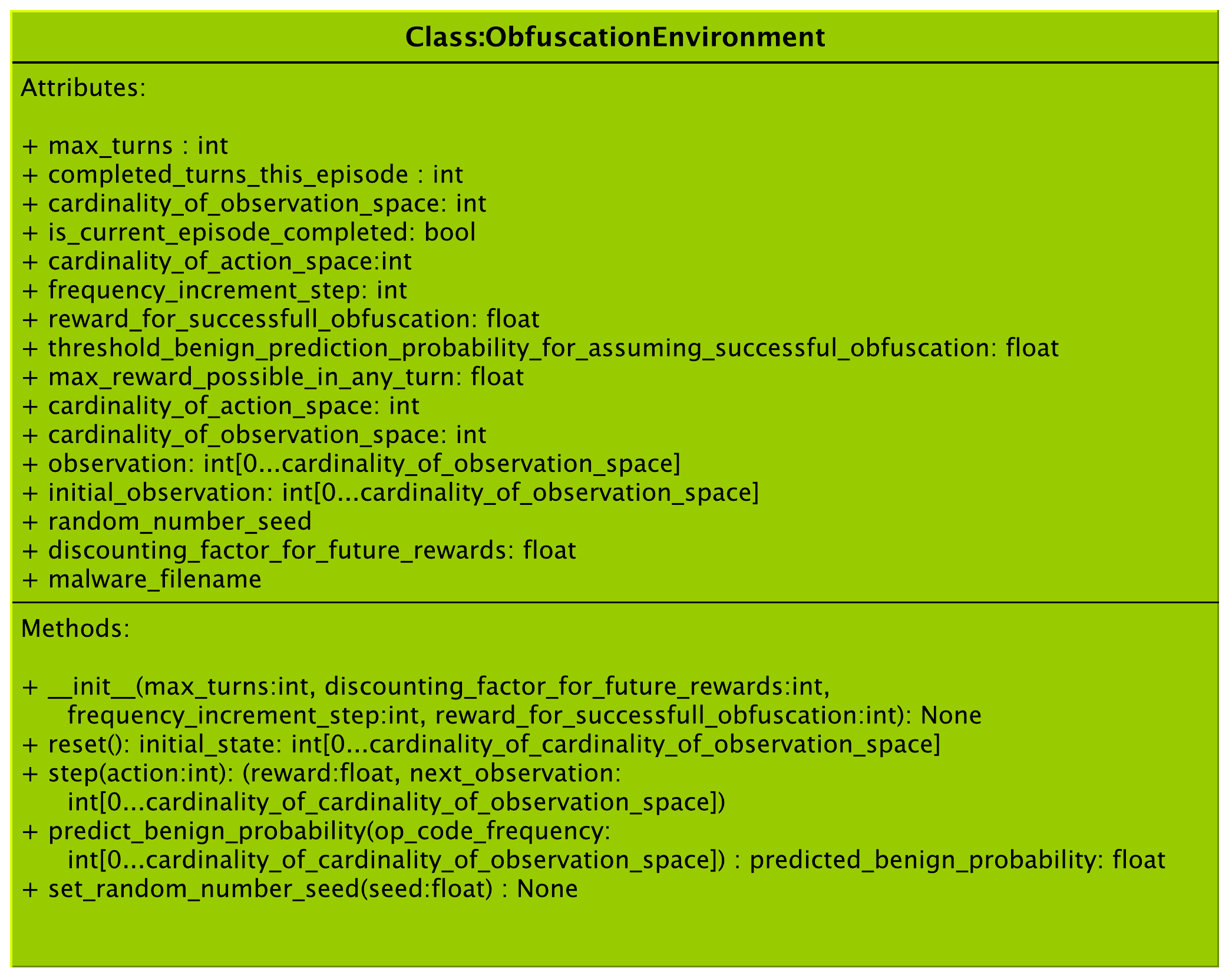}
    \caption{UML Diagram for the `ObfuscationEnvironment' Class}
    \label{fig:class_obfuscation_env}
\end{figure}

Combining the above three aspects of the environment as covered in the previous sub-sections, namely the `State' (\ref{env-state}), the `Action' (\ref{env-action}), and the `Reward Function' (\ref{env-reward}, we create a class for the malware obfuscation environment, the UML design of the same is given in fig \ref{fig:class_obfuscation_env}.

The two main role and corresponding methods of any custom environment are to reset an episode (using the `reset()' method) and to take the action suggested by the agent and in return deliver an appropriate `reward' and the subsequently changed state as the result of the action. This is done using the the `step(action)' action. The \_\_init\_\_ (python magic function to instantiate a class) instantiates the environment instance with custom parameters to train a unique DRL agent instance. Besides the environment parameters the agent's approximation function (DL model), and training configuration (epochs, batch size, weight initialization etc.) also determines the uniqueness of the thus trained agent. 
Besides these three methods for whose algorithm the pseudo-code is given below, there are other helper and utility methods defined in our custom environment implementation, but since these are not essentially required for a standard out-of-box reinforcement learning agent, we do not provide the pseudo-code of these here. The pseudo-code of the initialization method is given as \ref{alg:init}, for the reset method is given as \ref{alg:reset}, and for the step method is given as \ref{alg:step}.

\begin{algorithm}
    \caption{\_\_init\_\_() method} \label{alg:init}

    \begin{flushleft}
        \textbf{Input:}\newline
        $\mathbf{N_T}$ = max allowed turns in any episode: int, \newline
        $\mathbf{N_O}$ = cardinality of unique opcode set: int, \newline $\mathbf{Increment_{Step}}$ = any given opcode change step:int,\newline 
        $\mathbf{RF_{success}}$ = reward for successful completion of an episode: float, \newline
        $\mathbf{RF_{threshold}}$ = reward function parameter for minimum non-malicious probability threshold: float, \newline 
        $\mathbf{R_{Seed}}$ = random number seed for environment to instantiate the environment: int, \newline
        $\mathbf{\gamma}$ = discounting factor for future rewards: float \newline
        $\mathbf{Class_{MalwareDetect}}$ = Malware Prediction System Class: Object \newline
        \textbf{Output:}\newline 
        None
    \end{flushleft}
    
    \begin{algorithmic}
    \Procedure{init}{} \Comment{Initialize the Environment Class with appropriate values of the parameters to train a unique (DRL) agent instance} 
     \State $max\_turns \gets N_T$
     \State $cardinality\_of\_observation\_space \gets N_O$
     \State $cardinality\_of\_action\_space \gets N_O$
     \State $frequency\_increment\_step \gets Increment_{Step}$
     \State $reward\_for\_successful\_obfuscation \gets RF_{success}$
     \State $threshold\_non-malicious\_prediction\_probability \gets RF_{threshold}$
     \State $seed \gets R_{Seed}$
     \State $discounting\_factor \gets \gamma$
     \State $Function_{Predict} \gets \text{predict method of }Class_{MalwareDetect}$
     \State $is\_current\_episode\_completed \gets True$
     \State Set random number seed for implementation platform as seed
    \EndProcedure
    \end{algorithmic}
\end{algorithm}

\begin{algorithm}[htbp!]
\footnotesize
    \caption{reset() method} \label{alg:reset}
    \begin{flushleft}
        \textbf{Input:}\newline
            None \newline
        \textbf{Output:}\newline 
            array of opcode frequency: int[0...$N_O$]
    \begin{algorithmic}
    \Procedure{reset}{} \Comment{Resets the environment for a new episode to begin}
        \State Get the opcode of a random malware file from the malware repository
        \State $malware\_filename \gets \text{filename of the malware file}$
        % \State Convert the malware file to assembly code
        % \State Extract the opcode frequency from the assembly code
        \State $initial_observation \gets vector\_of\_opcode\_frequency$
        \State $turns\_completed \gets 0$
        \State $is\_current\_episode\_completed \gets False$
        \State $total\_episode\_reward \gets 0.0$ \Comment{used in performance evaluation}
        \State $total\_discounted\_episode\_reward \gets 0.0$ \Comment{for plotting and evaluation}
        \State return (initial\_observation)
    \EndProcedure
    \end{algorithmic}
    \end{flushleft}
\end{algorithm}
\raggedbottom

\begin{algorithm}[htbp!]
\footnotesize
    \caption{step() method} \label{alg:step}
    \begin{flushleft}
        \textbf{Input:}\newline
            $action\_index$ : int \newline
        \textbf{Output:}\newline 
            $reward$ : float \newline
            $new\_state$: int[0...$N_O$]
    \begin{algorithmic}
    \Procedure{step}{} \Comment{Step within an ongoing episode. Implements the action suggested by the agent and gives the corresponding reward and the new state to the agent.} 
        \If{is\_current\_episode\_completed = True}
            \State reset()
        \EndIf
        \State $new\_observation \gets current\_observation $
        \State $new\_observation[action\_index] \gets new\_observation[action\_index] + Increment_{Step} $
        \State $benign\_probability \gets Function_{Predict}(current\_observation)$
        \State $reward \gets benign\_probability - 0.5 $
        \State $total\_episode\_reward \gets \mathbb{\gamma} * total\_episode\_reward + reward$ 
            % \Comment{for performance evaluation}
        \State $total\_discounted\_episode\_reward \gets \mathbb{\gamma} * total\_discounted\_episode\_reward + reward$ 
            % \Comment{for plotting and evaluation}
        \State $total\_turns\_played \gets total\_turns\_played + 1$ \Comment{for plotting and evaluation}
        \State $current\_observation \gets new\_observation$
        \If{$benign\_probability \ge RF_{threshold} $} 
            \State{$ is\_current\_episode\_completed \gets True $} 
            \State store the new\_observation in obfuscated malware achieve
            \State compute similarity of new\_observation array with the initial\_observation array 
                % \Comment{for plotting and evaluation purpose}
        \EndIf
        \State return (reward, new\_observation)
    \EndProcedure
    \end{algorithmic}
    \end{flushleft}
\end{algorithm}
\raggedbottom

\section{Proximal Policy Optimization Agent Algorithm}\label{sec:rl-agent}
Given the constraints as covered in section \ref{env-action} earlier, we have a discrete action control requirement involving very high cardinality action space. Some of the most popular DRL agents for discrete action agents like the `Deep Q Networks' (DQN) \cite{DQN_Nature, DQN_Atari}, `Double DQN' (DDQN) \cite{Double_DQN}, and the `Dueling DQN' (DDQN) \cite{DQN_Nature} though could manage large state-space, but perform poorly for large/ continuous action space.

Deterministic Policy Gradient (DPG) \cite{DPG} based DRL approaches like the `Deep Deterministic Policy Gradient' (DDPG) \cite{DDPG} claims to deliver the best in class performance on RL tasks involving large, and even continuous action-space. The problem with such algorithms is that their line search based policy gradient update (used during optimization) either generates too big updates (which for updates involving non-linear trajectory makes the update go beyond the target) or makes the learning too slow. Since in the DRL paradigm non-linear gradients are quite common so algorithms based on line search gradient update does not prove very robust and cannot provide guarantees of near monotonic policy improvements. `Trust Region Policy Optimization' (TRPO) \cite{TRPO} an algorithm based on Trust-Region based policy update using `Minorize-Maximization' (MM) (second order) Gradient update claims to solve this problem and provide guarantee for near monotonic general (stochastic) policy improvement even for non-linear policies like that approximated by (deep) neural networks.

Additionally TRPO uses a mechanism called `Importance Sampling' to compute the expectancy of policy from previous trajectories instead of only the current trajectory to stabilize policy gradient. This method has an underlying assumption that the previous trajectory's distribution ($Q(x)$) is not very different from the current trajectory's distribution ($P(x)$). 

The policy gradient for a Stochastic Policy Gradient \cite{Stochastic_Policy_Gradient} method and associated algorithms like Actor Critic \cite{A3C} looks like below:
\begin{equation}
  \nabla_\theta (J_\theta) =  \mathbb{E}_{\tau \sim \pi_\theta (\tau)} [\nabla_\theta \log \pi_\theta(\tau) r(\tau)]  
  \label{eq:SPG}
\end{equation}
In equation \ref{eq:SPG}, the trajectory $\tau$ over which the samples for computing expectancy is gathered (to update the gradient $\nabla$ of the policy-value-function J), is also the same (current) trajectory of the policy $\pi$ (parameterized over $\theta$) as used in the update.

But in the case of TRPO using importance sampling and the past trajectory for sampling, this policy-value-function update looks as below equation \ref{eq:TRPO-PG}.
\begin{equation}
  \nabla_{\theta\prime} (J_\theta\prime) =  \mathbb{E}_{ \tau \sim \pi_\theta (\tau)} [\sum_{t=1}^T\nabla_\theta\prime \log \pi_\theta\prime (\prod_{t\prime=1}^t \frac{\pi_{\theta\prime}}{\pi_{\theta}}) (\sum_{t\prime=t}^T r)]  
    \label{eq:TRPO-PG}
\end{equation}
In the case of TRPO, the second order gradient update computation is very expensive, and hence for real size-able tasks it is seldom use. 

We use the PPO DRL agent algorithm (PPO) \cite{PPO}. PPO is an improvement over TRPO \cite{TRPO}. The PPO \cite{PPO} algorithm works similar to TRPO and is much easier to compute as it uses a linear-variant of the gradient update called the `Fisher Information Matrix' (FIM). In equation \ref{eq:TRPO-PG}, the trajectory is being sampled from the policy as it existed in previous time-step $(\pi_{\theta} = Q(x))$, whereas the expectancy over such collected samples are used to update the policy at next time-step $(\pi_{\theta\prime} = P(x))$. When the ratio of expectancy over the two trajectory distributions vary significantly $\frac{P(x)}{Q(x)}$ as in the case of linear gradient update in PPO, the previously stated assumption may not hold, leading to high variance in policy updates. To avoid this there are two methods that the PPO algorithm recommends. The first one use a `Adaptive KL Penalty' and the second one use `Objective Clipping'.

The recommended value of the  \textit{clipping-factor} ($\epsilon$) in the `Objective Clipping' is $\epsilon = 0.2$ \cite{PPO}. In the `Objective Clipping' mechanism if the probability ratio between the two trajectory's policies is not in the range $[(1-\epsilon),(1+\epsilon)]$ the `estimated advantage' is clipped. The DL function approximator that we use for the PPO algorithm's actor and critic network comprise of 2 hidden layers each, with each hidden layer having 64 neurons and a \textit{tanh} activation.

\section{Experimental Results}\label{sec:results}
Generating metamorphic malware using obfuscation as opcode level resulted in a high-cardinality action-space MDP, that the \textbf{PPO} algorithm based DRL agent was able to learn an efficient policy to solve. 

The resulting trained agents could obfuscate most of the malware and uplift their metamorphic ($\mathbb{P_{\text{non-malicious}}}$) probability of miss-classification error to a substantial degree to fail or evade even the best IDS which were even trained using the corresponding original malware variants.

\begin{figure}[htbp!]
    \centering
    \includegraphics[width=\linewidth, height=3.6in]{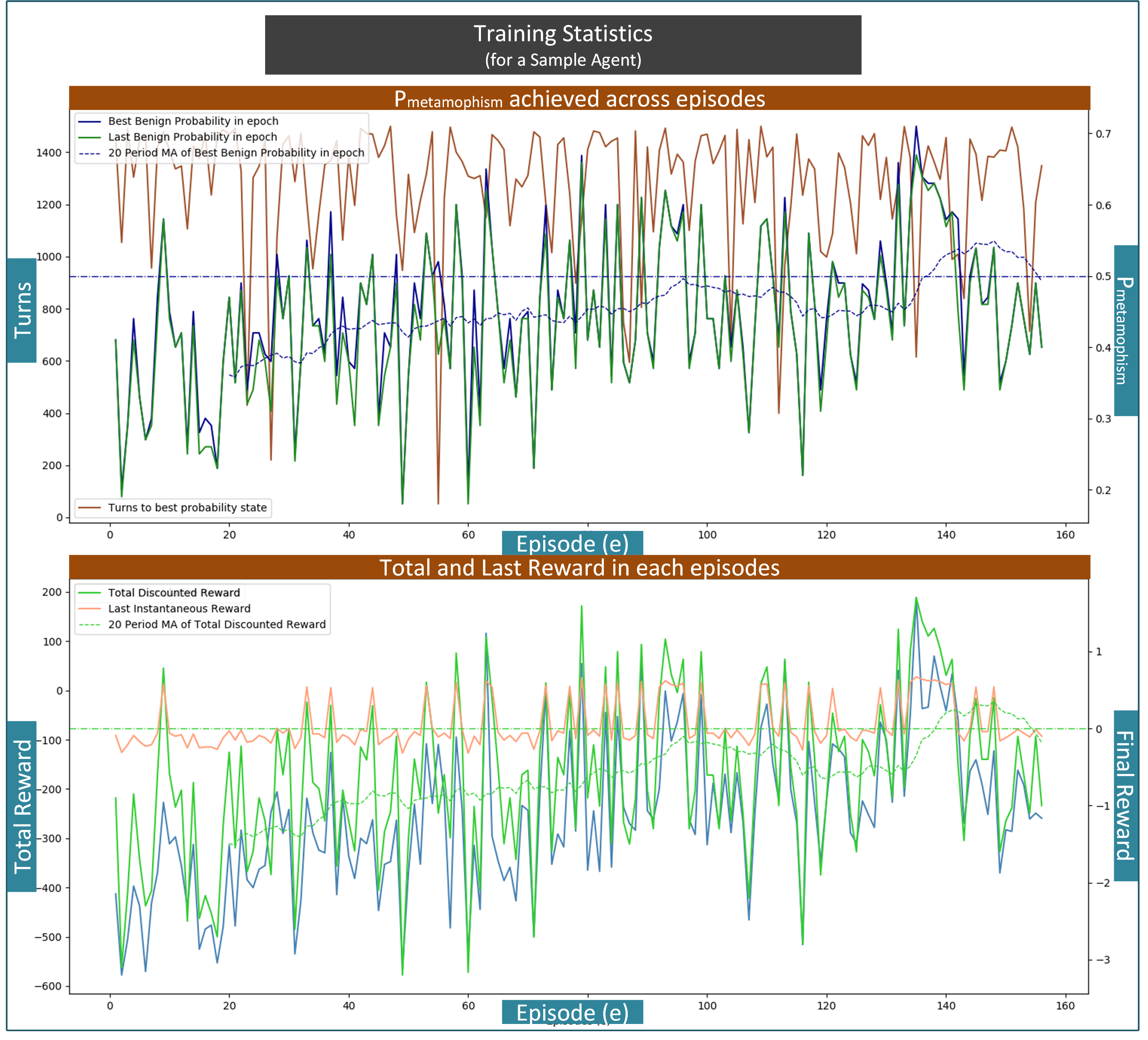}
    \caption{Training Statistics for a sample Agent}
    \label{fig:training_statistics_agent_1}
\end{figure}

\begin{figure}[ht]
    \centering
    \includegraphics[width=\linewidth, height=3.6in]{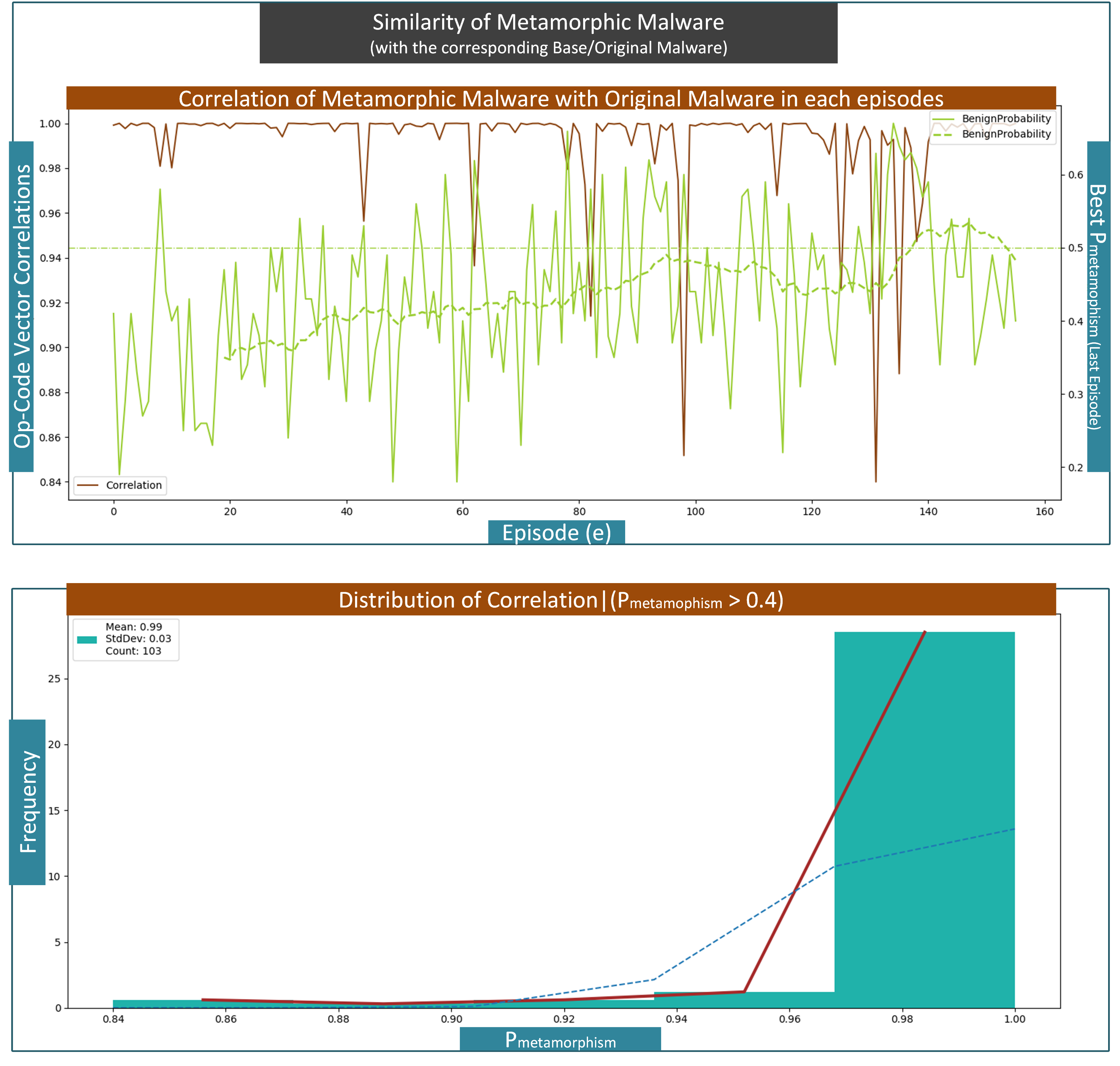}
    \caption{opcode Similarity for a sample Agent}
    \label{fig:opcode_similarity_agent_1}
\end{figure}

As shown in figure \ref{fig:training_statistics_agent_1}, the mean $\mathbb{P}_{NDMF}$ was uplifted by $\ge 0.45$  (from almost 0.0) indicating that the IDS could not effectively detect them as malicious unambiguously anymore. Also as shown in figure \ref{fig:opcode_similarity_agent_1} almost 1/3rd samples achieved a $\mathbb{P}_{NDMF} \ge 0.5$ (more likely to be non-malicious rather than malicious). The probability density function (PDF) distribution is left skewed, indicating that most of the samples were effectively obfuscated by the agent.

Another interesting observation is on the nature of opcode similarity between the original malware variants and their obfuscated version of the same. We use Pearson product-moment correlation coefficients between the opcode vectors to generate this similarity. The correlation is taken from the correlation matrix R, whose relationship with the co-variance matrix C, which is as given in the equation below:
\begin{equation}
    R_{ij} = \frac{ C_{ij} }{ \sqrt{C_{ii}*C_{jj} } } 
    \label{eq:pearson-correlation}
\end{equation}

\begin{figure}
    \centering
    \includegraphics[width=3.0in, height=4.2in]{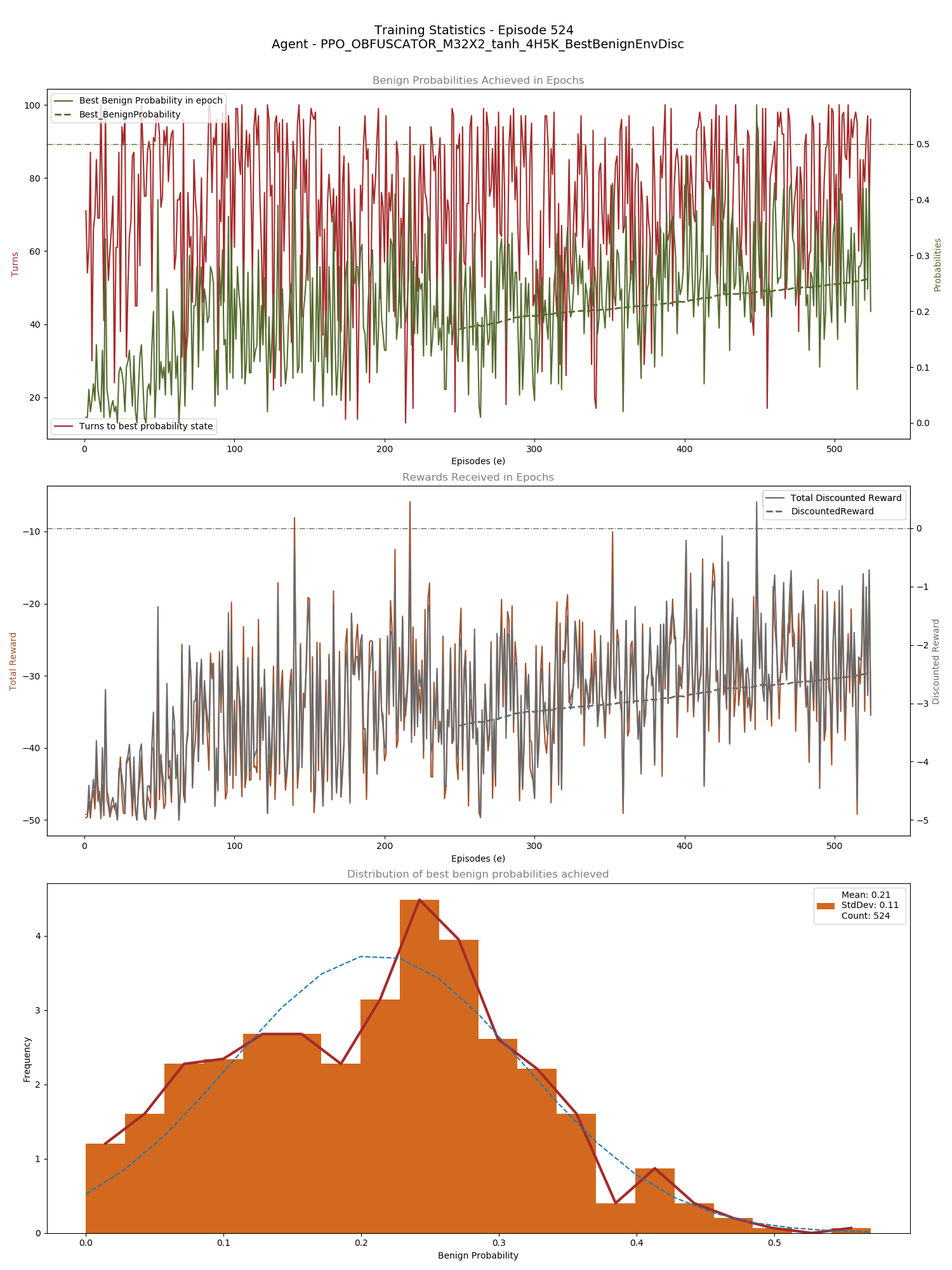}
    \caption{Training Statistics for Agent - 2}
    \label{fig:training_statistics_agent_2}
\end{figure}

In figure \ref{fig:opcode_similarity_agent_1}:
\begin{itemize}
    \item The first sub-plot shows that the resultant metamorphic instance's opcode frequency feature with that of the original non-obfuscated variant across episodes and also overlays the resulting $\mathbb{P}_{NDMF}$ of the resulting metamorphic vector.

    \item The second plot (histogram) indicates that the opcode frequency vector for the obfuscated variant is very similar to the original malware variant.
    
    \item The third plot is a similar histogram as sub-plot-2, but filtered for obfuscations that achieved a metamorphic probability $> 0.5$. It indicates that even for episodes where a high $\mathbb{P}_{NDMF}$ was achieved, the resulting opcode frequency vector is not very different from that of the original malware variant. 
\end{itemize}

\section{Discussion}\label{sec:discussion}
The insights from these observations are very critical as it indicates that with minimal efforts (adding additional opcode instructions using methods like junk-code insertion etc.) new obfuscations with very high metamorphic probability could be achieved with our system. Another important observation is that even with a net additive obfuscation the resulting metamorphic feature vector is very similar to that of the original and hence the resultant file-size (a function of total instructions) would also be very similar. This observation is important as many advanced IDS, besides featuring multiple classifiers, also has an unsupervised routing sub-component that helps improve the IDS' performance by routing the candidate file to the appropriate classifier based on the unsupervised sub-components routing. This routing sub-component could use ML techniques like clustering \cite{hemant_clustering}, or could even use simple heuristics based on file-size \cite{ashu_filesize}. If the file-size (for heuristic based) or opcode similarity (for clustering based) changes substantially then the resulting IDS may route the metamorphic variant to a different classier with which the evasion probabilities may not hold as high as indicated the lab settings. But having a high feature (and hence file-size) similarity, the resulting obfuscations from the A\underline{$\mathcal{D}$}VE\underline{$\mathcal{R}$}SARIA\underline{$\mathcal{L}$}uscator are much likely to be routed to the same classifier of an IDS, and hence the evasion probabilities would not be impacted.

\begin{figure}
    \centering
    \includegraphics[width=3.1in, height=4.2in]{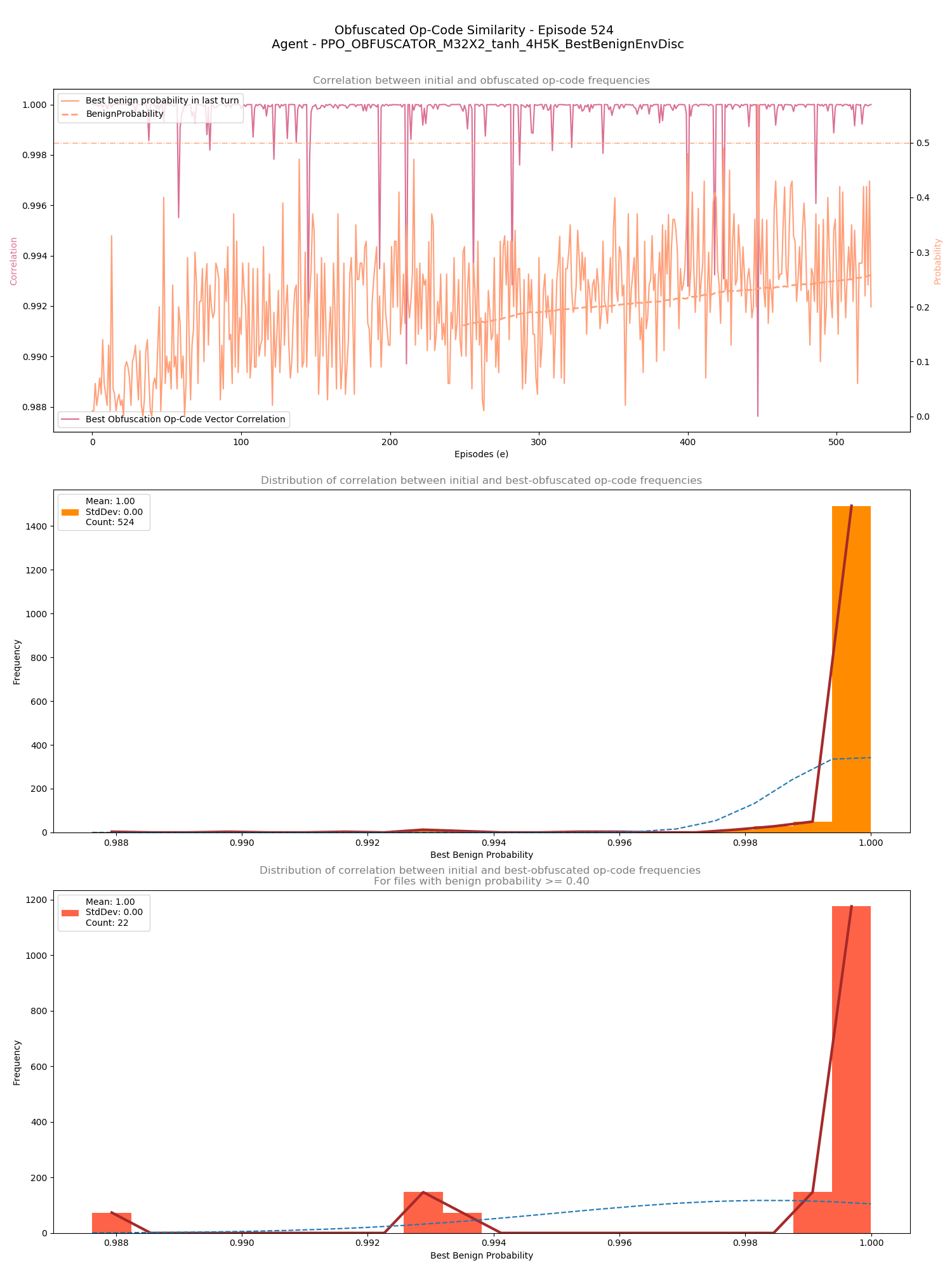}
    \caption{opcode Similarity for Agent - 2}
    \label{fig:opcode_similarity_agent_2}
\end{figure}

\section{Conclusion}\label{sec:conclusion}

We designed and developed a novel system that uses specialized (adversarial) deep reinforcement learning to obfuscate malware at the opcode level and create multiple metamorphic instances of the same and named it ADVERSARIALuscator. To the best of our knowledge, A\underline{$\mathcal{D}$}VE\underline{$\mathcal{R}$}SARIA\underline{$\mathcal{L}$}uscator is the first-ever system that adopts the Markov Decision Process based approach to convert and find a solution to the problem of creating individual obfuscations at the opcode level. This is important as the machine language level is the least at which functionality could be preserved so as to mimic an actual attack effectively. A\underline{$\mathcal{D}$}VE\underline{$\mathcal{R}$}SARIA\underline{$\mathcal{L}$}uscator is also the first-ever system to use efficient continuous action control capable deep reinforcement learning
Experimental results indicate that A\underline{$\mathcal{D}$}VE\underline{$\mathcal{R}$}SARIA\underline{$\mathcal{L}$}uscator could raise the metamorphic probability of a corpus of malware by $\ge 0.45$. Additionally, more than $33\%$ of metamorphic instances generated by A\underline{$\mathcal{D}$}VE\underline{$\mathcal{R}$}SARIA\underline{$\mathcal{L}$}uscator were able to evade even the most potent IDS and penetrate the target system, even when the defending IDS could detect the original malware instance. 
Hence, A\underline{$\mathcal{D}$}VE\underline{$\mathcal{R}$}SARIA\underline{$\mathcal{L}$}uscator could be used to generate a swarm of very potent metamorphic malware. Such an advanced attack could effectively mimic an extreme `zero-day attack’ from multiple simultaneous metamorphic malware and hence can create ideal data required to bolster the defences of an IDS against an actual and potential AI based malware attack. 
However, we believe that incorporating some of the very recent developments in Multi-Agent Reinforcement Learning (MARL) can lead to the development of an even more potent system with capabilities of coordinated swarms of malware, and the work in this direction is in progress.

\bibliographystyle{unsrt}
\bibliography{main}

\end{document}